\documentclass[twocolumn,floatfix,superscriptaddress,showpacs,showkeys]{revtex4}
\usepackage{graphicx}
\usepackage{amsmath}
\usepackage{booktabs}
\usepackage{amssymb}
\usepackage{multirow}
\usepackage{makecell}
\usepackage{textcomp}
\usepackage{subfigure}
\usepackage{phonetic}
\usepackage{extarrows}
\usepackage{color}
\usepackage{float}
\usepackage[colorlinks,citecolor=blue,linkcolor=blue]{hyperref}
\usepackage[utf8]{inputenc}

\begin{document}

\title{Pseudoscalar Yukawa Coupling Induced Helical Asymmetry in Fermionic Preheating} 

\author{Ke FU}
\affiliation{College of Physics and Electronic Information, Inner Mongolia Normal University, 81 Zhaowuda Road, Hohhot, 010022, Inner Mongolia, China}
\affiliation{Inner Mongolia Key Laboratory of Applied Condensed Matter Physics, Inner Mongolia Normal University, 81 Zhaowuda Road, Hohhot, 010022, Inner Mongolia, China}
\author{Li-Shuang Liu}
\affiliation{College of Physics and Electronic Information, Inner Mongolia Normal University, 81 Zhaowuda Road, Hohhot, 010022, Inner Mongolia, China}
\affiliation{Department of Physics, College of Science, Yanbian University, Yanji, Jilin 133002, China}
\author{Yu-Feng Wang}
\affiliation{College of Physics and Electronic Information, Inner Mongolia Normal University, 81 Zhaowuda Road, Hohhot, 010022, Inner Mongolia, China}
\affiliation{School of Information and Optoelectronic Science and Engineering, South China Normal University, Guangzhou, 510006, China}
\author{Xin-Yu Gu}
\affiliation{College of Physics and Electronic Information, Inner Mongolia Normal University, 81 Zhaowuda Road, Hohhot, 010022, Inner Mongolia, China}
\affiliation{Inner Mongolia Key Laboratory of Applied Condensed Matter Physics, Inner Mongolia Normal University, 81 Zhaowuda Road, Hohhot, 010022, Inner Mongolia, China}
\author{Xi-Bin Li}
\email{lxbimnu@imnu.edu.cn}
\affiliation{College of Physics and Electronic Information, Inner Mongolia Normal University, 81 Zhaowuda Road, Hohhot, 010022, Inner Mongolia, China}
\affiliation{Inner Mongolia Key Laboratory of Applied Condensed Matter Physics, Inner Mongolia Normal University, 81 Zhaowuda Road, Hohhot, 010022, Inner Mongolia, China}

\begin{abstract}
In this study, we investigate the fermionic helical asymmetry during preheating by introducing a model in which the Dirac field $\psi$ couples to
the scalar inflaton $\phi$ directly through the pseudoscalar Yukawa mechanism described by the term $- g\frac{\phi}{f} m_\psi \bar{\psi}\mathrm{i}\gamma^5 \psi$.
Following inflation, as the inflaton oscillates around the minimum of its potential, the effective frequency of Dirac field changes non-adiabatically, resulting in particle production.
These oscillations also indicate that the pseudo-mass $m_5 = g \phi m_\psi/f$ changes the sign, leading to the production of helical-asymmetric fermions.
We employ the WKB method to analytically calculate the transition relations of the Bogoliubov coefficients,
which is valid for an arbitrary number of productions and demonstrates a significant difference compared to the case without asymmetry.
In the expanding universe, the presence of pseudo-mass $m_5$ alters the Gaussian distributions for each helical state, 
since the combined effects of Pauli blocking and parametric resonance produce coherent enhancement and coherent suppression in certain frequency ranges. 
This departure arises from the imaginary component of the first-order adiabatic complex phase, denoted as $\Theta^{(r)}_{1}$.
The real part of $\Theta^{(r)}_{1}$ significantly enhances the number density of each helical state after eight oscillations of the inflaton since the coherent superposition.
These conclusions have also been validated by the numerical results.
\end{abstract}
\maketitle

\section{Introduction}
\label{intro}
In the standard cosmological model, an epoch is believed to occur between the inflationary period and the radiation-dominated era of the Universe.
During this epoch, a mechanism exists to convert the inflaton energy density, following the end of inflation, into radiation energy density.
Depending on whether the mechanism is perturbative or non-perturbative, this epoch is referred to as reheating \cite{reheating1} or preheating \cite{preheating1}, respectively.
Preheating of bosons \cite{preheating_boson1,preheating_boson2,preheating_boson3,preheating_boson4} is characterized by a highly efficient and explosive particle production, resulting from the coherent oscillations of the inflaton field.
This permits substantial production even when the decay of a single particle is kinematically forbidden. 
It has been noticed \cite{preheating_fermion1,preheating_fermion2,preheating_fermion3,preheating_fermion4,preheating_fermion5,preheating_fermion6}
that preheating of fermions can also be very efficient, despite the production in this case being limited by Pauli blocking.
All the scenarios proposed thus far have assumed preheating at higher energy scales, predicting the existence of new particles that are heavier than the Standard Model (SM) particles.
The QCD preheating recently proposed in the literature \cite{QCD1,QCD2} exhibits low-scale behaviour characterised by a dynamic and non-adiabatic baryon chemical potential.

It is widely recognized that fermions can be generated through Yukawa interactions with oscillating scalar fields following the period of inflation \cite{preheating_fermion1}.
In certain models, preheating is predicted to create fermions with masses that can be considerably greater than that of the inflaton \cite{preheating_fermion5}.
This mechanism could play a crucial role in leptogenesis scenarios, where the generation of baryon asymmetry occurs through the decay of right-handed neutrinos
with masses close to the grand unified theory scale \cite{GUT1,GUT2,GUT3}.
Reheating also plays a crucial transitional role that determines when and how the rotating axion-like particle transitions from its rotation phase (responsible for baryogenesis) 
to the coherent oscillation phase that produces cold dark matter \cite{bary1}.
Further, Ref. \cite{lepto1} examined the production of neutrinos during the relaxation and oscillations of the Higgs condensate
after inflation as a potential mechanism for generating matter-antimatter asymmetry through leptogenesis.
The cosmological implications of heavy fermionic particles being gravitationally created during the inflationary epoch have long been recognized and thoroughly studied \cite{fermion1,fermion2,fermion3}.
Gravitational particle production has also been explored as a viable channel for generating super-heavy fermions in the inflationary era \cite{massive1,massive2,massive3}.
Meanwhile, the dynamics of fermions coupled to axions in the post-inflationary universe were initially investigated within the framework of spontaneous baryogenesis \cite{Baryo1,Baryo2,Baryo3,Baryo4},
and this idea has recently been revisited and adapted to scenarios of thermal and non-thermal leptogenesis \cite{lepto2,lepto3,lepto4,lepto5}.    
Even for moderately small Yukawa couplings $y \gtrsim 10^{-8}$, parametric resonance, kinematic blocking, 
and Pauli suppression strongly affect fermion production and cannot be ignored \cite{asymmetry3}. 
In the context of asymmetric reheating \cite{asymmetry4}, an appropriate choice of Yukawa couplings can lead to the thermalization of the dark-matter candidates in the hidden sector, 
provided unitarity bounds are requires and the hidden sector itself reaches thermal equilibrium.

WKB (Wentzel-Kramers-Brillouin) approximation is commonly employed in semiclassical calculations in quantum mechanics, where the wave function is expressed as an exponential function.
In this approach, either the amplitude or the phase is assumed to vary slowly \cite{WKB1}.
This method has been applied to pair production in a vacuum induced by an alternating field \cite{WKB2}.
The Schwinger mechanism \cite{WKB3} can be analyzed using the complex WKB method and remains an active area of research in inflationary cosmology \cite{WKB4,WKB5}.
For cosmological preheating, a similar calculation is presented in Ref. \cite{WKB6}.
In Refs. \cite{WKB7,WKB8}, the method has been used to calculate the effects of higher-dimensional interactions during preheating.
The exact WKB analysis \cite{WKB9,WKB10} is employed into the study of particle formation involving certain exotic interactions \cite{WKB11}.

Couplings between vector gauge fields and rolling pseudo-scalar fields lead to a significant amplification of the gauge field polarizations during de Sitter expansion \cite{pseudoscalar3,pseudoscalar4,pseudoscalar5};
however, the impact of such couplings on fermion fields has been scarcely investigated.
Reference \cite{axion1} claims that, during preheating, the pseudoscalar-axion oscillates and fermions with both helicities are produced asymmetrically,
resulting in unequal number-densities of left- and right-helicity fermions.  
Preheating in the presence of fermions has been systematically investigated in the models involved to mechanics that break helical symmetry. 
Related works have also obtained the baryon asymmetry via Schwinger effect in axion inflation \cite{asymmetry1}, hydrodynamical approaches to chirality production of magnetogenesis \cite{asymmetry2}, 
fermion reheating with quartic potentials \cite{asymmetry3}, and other aspects of low-scale preheating dynamics \cite{asymmetry4,asymmetry5}. 
The present analytical and numerical framework provides a complementary method for studying helical asymmetries in similar models. 

The WKB approximation has also been employed to compute the helicity asymmetry of fermions, as the standard framework for analyzing non-perturbative particle production.
In this work, we introduce the pseudoscalar coupling $g \frac{\phi}{f} m_\psi \bar{\psi} \mathrm{i} \gamma^5 \psi$, which interacts between the scalar inflaton $\phi$ and the Dirac fermion $\psi$.
Furthermore, we apply the WKB approximation to study a fermionic preheating model incorporating the same Yukawa coupling to break the helical symmetry. 
Within this framework, the Bogoliubov coefficients are computed analytically, enabling a more detailed examination of the helical production process.  
We assume that the preheating phase primarily generates the fermionic helical asymmetry through non-adiabatic oscillations around the minimum of the inflaton potential.
This assumption is also based on the consideration that the inflation phase produces the same helicities.
The primary aim is to derive analytical expressions for the Bogoliubov coefficients of Dirac fields after each oscillation, focusing particularly on the coefficients $\beta^{(r)}_n$,
which depend on the helical states $r$ arising from the pseudoscalar coupling.

Although Ref.~\cite{axion1} has studied pseudoscalar axion inflation both during and after inflation by introducing the axion-fermion interaction $\frac{g}{f}\partial_\mu\phi\,\bar{\psi}\gamma^\mu\psi$, 
the present work provides several important supplementary results.
We confirm that, as the axion oscillates after inflation, the produced fermions can develop a significant helicity asymmetry.
Using this method, we analytically compute the Bogoliubov coefficients and obtain more details about the helical fermion production, as results that has not been addressed before.
A larger fermion mass further enhances the production of this helical asymmetry.
The major improvement of this work is based on the derivation of complete analytical results for pseudoscalar Yukawa preheating. 
In particular, we show that fermionic helical asymmetry is generated only in certain ranges where the comoving momentum $k$ is much smaller than the characteristic wavelength $(m_\psi a')^{-1/2}$, i.e., $k<(m_\psi a')^{-1/2}$. 
Once this comoving wavelength enters the characteristic horizon, the helical asymmetry is dramatically suppressed.
Mathematically, the axion-fermion pseudoscalar interaction introduces an imaginary contribution to the WKB phase that depends on the helicity state $r$, which is responsible for the generation of helical asymmetry. 
Meanwhile, the real part of the WKB phase governs the coherent superposition of the modes after several oscillations. 
These important features were not addressed in the previous work.

This paper is organized as follows. In Sec. \ref{analy}, we introduce the pseudoscalar coupling at preheating phase and derive analytical results for the Bogoliubov coefficients of helical Dirac fields.
Based on these results, we calculate the quasi-analytical expressions for the fermionic densities of each helical state in Sec. \ref{number}.
Then, in Sec. \ref{Num}, the numerical results are presented for both static and expanding universes, and some interesting phenomena arising from helical asymmetry are further discussed.
Finally, in Sec. \ref{conclusion}, we provide a brief conclusion of this work.

\section{Pseudoscalar mechanism and solutions at preheating \label{analy}}
\subsection{The model}

The scalar field that drives the exponential expansion can decay into various fundamental particles after the inflationary era.
In this study, we examine the production of fermions during the preheating era via a pseudoscalar Yukawa coupling, in which the fermions are directly coupled to the scalar field to break symmetry.
The Lagrangian is expressed as follows \cite{pseudoscalar1,pseudoscalar2}:
\begin{align}
    \mathcal{L}/\sqrt{-g} = \frac{1}{2} \partial_\mu  \phi \partial^\mu \phi & - V(\phi)+\mathrm{i} \bar{\psi} \gamma^\mu D_\mu \psi  \nonumber\\
    &  -m_\psi \bar{\psi} \psi - g\frac{\phi}{f} m_\psi \bar{\psi}\mathrm{i}\gamma^5 \psi, \label{L}
\end{align}
where $\phi$ denotes the single scalar field; $\psi$ and $\bar\psi=\psi^\dagger \gamma^0$ represent the Dirac spinor field and its adjoint,
respectively, which are coupled to the scalar field via the axion-fermion mechanism $g(\phi/f) m_\psi \bar{\psi}\mathrm{i}  \gamma^5 \psi$;
and $m_\psi$ is the static mass of the fermion.
The line element for Friedmann-Robertson-Walker metric is given by
\begin{align}
    \mathrm{d}s^{2}&=\mathrm{d}t^2-a^2\delta_{ij}\mathrm{d}x^i\mathrm{d}x^j \nonumber \\
    &=a^2{\left(\mathrm{d}\tau^2-\delta_{ij}\mathrm{d}x^i\mathrm{d}x^j\right)},  \label{metric}
\end{align}
with $\mathrm{d}t=a\mathrm{d}\tau $, where $a$ is the cosmic scalar factor, $t$ is the cosmic time, and $\tau$ is the conformal time.
The gamma matrices in Friedmann-Robertson-Walker universe are expressed as follows %(with the sign convention for the metric $(+\ -\ -\ -)$)
\begin{align}
    \bar{\gamma}^0=\gamma^0,\quad \bar{\gamma}^i=\frac1a\gamma^i,
\end{align}
where $\gamma^\mu$ denotes the gamma matrices in flat spacetime.
The derivative operator in the Lagrangian \eqref{L} is expressed as \cite{Dirac1,Dirac2}
\begin{align}
    D_\mu\psi=\partial_\mu\psi+\Gamma_\mu\psi, \label{D}
\end{align}
with the spin connections in curved spacetime defined as $\Gamma_\mu$, given by $\Gamma_0=0,\ \Gamma_i=\frac{\dot{a}}{2}\gamma^0\gamma^i$\cite{Dirac3}.
The spatial contraction of these connections reads
\begin{align}
    \bar{\gamma}^i\Gamma_i=\frac32H\gamma^0, \label{contraction}
\end{align}
where $H = \mathrm{d} \ln a/\mathrm{d}t$ represents the cosmic Hubble parameter.

We next study the fermion production after inflation, that is while the inflaton field coherently oscillates about the minimum of the potential
\begin{align}
    V(\phi)=\frac{1}{2}m_{\phi}^{2}\phi^{2},\quad m_{\phi}\simeq3\times10^{13}\mathrm{GeV}. \label{V}
\end{align}
Therefore, after few oscillations, the scalar field evolves according to the formula
\begin{align}
    \phi(t)\simeq\frac{M_p}{\sqrt{3\pi}}\frac{\sin{(m_\phi t)}}{m_\phi t}. \label{phi}
\end{align}

It is worth noting that the potential in Eq.~\eqref{V} represents the typical form used in chaotic inflation. 
However, this is not the specific model we investigate in detail here.
In this work, we do not explore the explicit form of the inflaton potential. 
Instead, we focus on their phenomenological consequences during the preheating phase. 
To allow direct comparison with the existing preheating literature, we adopt the widely used quadratic potential $V(\phi) = \frac{1}{2} m_\phi^2 \phi^2$ as our precondition. 
This enables to sufficiently illustrate the effects of the additional features on the preheating dynamics \cite{quadratic}.
Preheating in many observationally favored inflationary models, such as Starobinsky inflation and $\alpha$-attractor models, can be well approximated by a quadratic potential near the minimum. 
Therefore, we only focus the potential with quadratic $\frac12 m_\phi^2 \phi^2$ form around the its minimum. 
In addition, previous studies of fermionic preheating have generally assumed that the process occurs at a high energy scale, 
which is also fixed by the amplitude of the scalar perturbations inferred from CMB measurements \cite{CMB}. 
Consequently, the inflaton mass is typically set to $m_\phi \simeq 3 \times 10^{13}\, \mathrm{GeV}$ to be consistent with this assumption.
However, this choice is not essential for our following analysis, since the fermionic production rate is independent of this parameter, which will be analytically presented in the following sections. 

\subsection{Equations about helical fermions}

The Dirac-like equation for the fermion field is derived by varying the action with respect to the field $\psi$:
\begin{align}
    \left( \frac{\mathrm{i} }{a} \gamma^\mu \partial_\mu + \mathrm{i}  \frac{3}{2} H\gamma^0 - m_\psi - \mathrm{i} g \frac{\phi}{f} m_\psi \gamma^5 \right) \psi = 0. \label{psi}
\end{align}
Here the partial operator is about the conformal time, $\partial_\mu\equiv(\partial_\tau,\partial_i)$.
We now introduce the canonically normalized fermion fields $u$ and $v$ in momentum space to rewrite the Dirac field as
\begin{align}
    \psi(x,\tau)=\int\frac{\mathrm d^3k}{(2\pi a)^{3/2}}\mathrm e^{\mathrm i\mathbf{k}\cdot\mathbf{x}}\sum_{r=\pm}\left[u_r(k,\tau)\hat b_r(\mathbf k)+ \right. \nonumber\\
    \left.v_r(k,\tau)\hat d_r^\dagger(-\mathbf k)\right]. \label{uv}
\end{align}
where $\hat b_r$ and $\hat d_r^\dagger$ are the annihilation and creation operators for particles and antiparticles, respectively, with helical state $r$.
Substituting this into the Dirac-like equation \eqref{psi}, it acquires the equation with a more familiar form:
\begin{align}
    \left(\mathrm{i} \gamma^0 \partial_\tau -k_i \gamma^i - m_\psi a - \mathrm{i} m_5 a \gamma^5\right) u_r = 0. \label{u}
\end{align}
with pseudo-mass
\begin{align}
    m_5 = g \phi m_\psi / f.  \label{m5}
\end{align}

We now choose the gamma matrices as
\begin{align}
    \gamma^0 = \begin{pmatrix} I & 0 \\ 0 & -I \end{pmatrix}, \
    \gamma^i = \begin{pmatrix} 0 & \sigma^i \\ -\sigma^i & 0 \end{pmatrix}, \
    \gamma^5 = \begin{pmatrix} 0 & I \\ I & 0 \end{pmatrix},  \label{gamma_matrices}
\end{align}
where $I=I_{2\times2}$ is the $2\times2$ identity matrix and $\sigma^i$ are the general Pauli matrices.
It is convenient to take the momentum along the third direction, $k = k_z$, and to define the Dirac field in terms of helical states:
\begin{align}
    &u_r = \left[u_{r,+}(\tau) \psi_r, u_{r,-}(\tau) \psi_r\right]^T, \nonumber\\
    &v_r = \left[v_{r,+}(\tau) \psi_r, v_{r,-}(\tau) \psi_r\right]^T, \label{ur}
\end{align}
where $\psi_+ = (1, 0)^T$ and $\psi_- = (0, 1)^T$ denote the eigenvectors of the helicity operator $\mathbf \sigma\cdot\mathbf k/|\mathbf k|$, with eigenvalues $r=\pm1$. 
Then applying Eqs. \eqref{gamma_matrices} and \eqref{ur}, the Dirac-like equation \eqref{u} rewrites
\begin{subequations}
\begin{align}
    \mathrm{i} u'_{r,+} - m_\psi a u_{r,+} = (r k + \mathrm{i} m_5 a) u_{r,-}, \label{up}\\
    \mathrm{i} u'_{r,-} + m_\psi a u_{r,-} = (r k - \mathrm{i} m_5 a) u_{r,+}, \label{un}
\end{align}
\end{subequations}
with prime $'$ representing the partial derivative with respect to the conformal time $\tau$.
The equation for $u_{r,+}$ is derived  by multiplying both sides of Eq. \eqref{up} by $\frac{r k - \mathrm{i} m_5 a}{k^2 + m_5^2 a^2}$ and then substituting the result into Eq. \eqref{un}.
Correspondingly, $u_{r,-}$ is obtained by multiplying both sides of Eq. \eqref{un} by $\frac{r k + \mathrm{i} m_5 a}{k^2 + m_5^2 a^2}$ and then substituting the result into Eq. \eqref{up}.
Thus, we have the equations for $u_{r,\pm}$:
\begin{align}
    &\left[\frac{\mathrm d^2}{\mathrm d \tau^2} + \left(\ln \frac{1}{r k\mp\mathrm{i} m_5 a} \right)' \frac{\mathrm d}{\mathrm d \tau} +\right. \nonumber\\
    &\quad\quad\quad\quad k^2 \pm \mathrm{i} (m_\psi a)' + m^2_\psi a^2 + m_5^2 a^2 \Bigg] u_{r,\pm} = 0. \label{upm}
\end{align}
This equation indicates that the presence of the pseudoscalar coupling in Eq. \eqref{L} breaks 
the symmetry of $u_{r,\pm}$ with respect to the helical state $r$.
This clearly differs from the case in the absence of such term \cite{preheating_fermion1}.
Without the pseudoscalar coupling to the inflaton, fermion production is helicity-symmetric, as anticipated.
Once the coupling is included, the production rates for the two helicity states become asymmetric.

We stress that helicity and chirality are distinct concepts in particle physics.
For massless or ultra-relativistic particles, however, the two notions coincide and describe the same physical state.
In the present work, we focus on the preheating stage taking place during the radiation-dominated era, where fermions still possess their ordinary mass and must be treated as non-relativistic.
Consequently, it is helicity, rather than chirality, that serves as the proper quantum number to characterize the fermionic asymmetry.
Particularly, helicity-asymmetric neutrinos may generate the matter-antimatter asymmetry through leptogenesis scenarios.  

\subsection{Evolution equations about helical fermions}

We now utilize the fact that, in the presence of a pseudo-mass $m_5=g \phi m_\psi/f$, particle creation with a specific helicity occurs only during very short intervals around the points where $\phi_* = 0$.
At these points, the pseudoscalar term $m_5\bar\psi\mathrm i\gamma^5\psi$ vanishes.
This consideration enables the application of the same formalism used in the bosonic \cite{Neq} and helical-symmetric fermionic \cite{preheating_fermion1} cases to the production of asymmetric fermions.
While the Bogolyubov coefficients can be treated as constant when far from the zeros of the pseudo-mass $m_5$, a sudden variation occurs whenever $\phi$ crosses $\phi_* = 0$.
Since the production interval is very narrow, one can safely neglect the expansion of the universe during this period and approximate the function linearly as $\phi(\tau) \approx \phi_* + \phi'(\tau_*)(\tau - \tau_*)$,
where $\tau_*$ is the moment so that $\phi(\tau_*) = 0$.
%It is notable that the parameters $\tau_{*}$ and $\phi_*$ are dependent on the oscillation times $n$.
The oscillatory mode of the scalar field, as expressed in Eq. \eqref{phi}, indicates that fermion production occurs infinitely many times.
As a consequence, the primary question is to test the convergence of the total fermion density in the presence of infinite oscillations, which will be examined theoretically in Sec. \ref{semi}.

We now consider the evolution equation \eqref{upm} near the point $\tau_{*n}$, which corresponds to the moment when $\phi=\phi_*\equiv0$ for the $n$-th time (i.e., the \(n\)-th fermionic production event).
Since the pseudo-mass $m_5$ vanishes, fermionic production is confined to a very short interval, specifically $m_\phi^{-1} \ll H^{-1}$.
One can neglect the expansion of the universe during this period and express the equation for $\phi(\tau)$ in a linearized form.
We thus write
\begin{align}
    m_\psi a(\tau) \approx m_\psi a'_{*n} (\tau - \tau_{*n}),\quad a_{*n}\equiv a(\tau_{*n}), \nonumber\\
    a\phi(\tau) \approx a_{*n} \phi'_{*n} (\tau - \tau_{*n}), \quad \phi'_{*n}\equiv\frac{\mathrm d\phi}{\mathrm  d\tau}\Big|_{\tau_{*n}}. \label{para1}
\end{align}
Then define the parameters as follows
\begin{align}
    &p_n = k/\sqrt{m_\psi a'_{*n}}, \ z = - (m_\psi a'_{*n})^{1/2} (\tau - \tau_{*n}), \nonumber\\
    &\xi_n = \frac{g\phi'_{*n} a_{*n}}{f a'_{*n}} = \frac{g\phi'_{*n}}{f \mathcal H_{*n}}, \quad  \mathcal H_{*n}\equiv\frac{a'_{*n}}{a_{*n}}, \label{para2}
\end{align}
and insert them into the evolution equation \eqref{upm}; thus it yields
\begin{align}
    &\frac{\mathrm d^2 u_{r,\pm}}{\mathrm d z^2} + \frac{1}{z \pm \mathrm i r p_n / \xi_n} \frac{\mathrm d u_{r,\pm}}{\mathrm d z} + \nonumber\\
    &\quad\quad\quad\quad \left[ p_n^2 \pm \mathrm{i} + (1+\xi_n^2)z^2 \right] u_{r,\pm} = 0. \label{u_z}
\end{align}
For the sake of convenience, we will omit the subscript $n$ in the following equations.
We now introduce the new variable $\tilde u_{r,\pm} = \sqrt{z \pm \mathrm i r p / \xi} u_{r,\pm}$ to transform Eq. \eqref{u_z} into
\begin{align}
    &\frac{\mathrm d^2 \tilde{u}_{r,\pm}}{\mathrm d z^2} + %\nonumber\\
    \left[ p^2 \pm  \mathrm{i} + (1+\xi^2)z^2 + \frac{1}{4(z \pm \mathrm{i}r p/\xi)^2} \right] \tilde{u}_{r,\pm}\nonumber\\
    &\quad\quad =0. \label{evol}
\end{align}
It is seen shortly later that the term about $(z \pm \mathrm{i}r p/\xi)^{-2}$ in the equation is the primary factor responsible for generating helical asymmetry.

\subsection{Analytical evaluation of the occupation number}

The derivation of the analytical formulas using the WKB method involves considering asymptotic solutions of Eq. \eqref{evol}.
We seek solutions valid for $\phi$ far from $\phi_*=0$, where the adiabatic condition $\mathrm d\omega_\pm/\mathrm d z\ll\omega_\pm^2$ holds
as long as $z\gg(1+\xi^2)^{-1/4}\geq p/(1+\xi^2)^{1/2}$, with
\begin{align}
    \omega_\pm^2(z) = p^2 + (1+\xi^2)z^2 + \frac{1}{4(z \pm  \mathrm{i} r p/\xi)^2}.
\end{align}
In this regime, the most general solutions of Eq. \eqref{evol} take the form of an exponential complex integral:
\begin{subequations} \label{uB}
\begin{align}
    &\tilde{u}_{r,+}(z)= \nonumber\\ &\quad \alpha^{(r)} \frac{\mathrm{e}^{\mathrm{i} \int^z \omega_+(z') \mathrm{d}z'}}{\sqrt{2 \omega_+(z)}} +
        \beta^{(\bar{r})} \frac{\mathrm{e}^{-\mathrm{i} \int^z \omega_+(z') \mathrm{d}z'}}{\sqrt{2 \omega_+(z)}}, \label{u+} \\
    &\tilde{u}_{r,-}(z)= \nonumber\\ &\quad  -\beta^{(\bar{r})*} \frac{\mathrm{e}^{-\mathrm{i} \int^z \omega_-(z') \mathrm{d}z'}}{\sqrt{2 \omega_-(z)}}
        +\alpha^{(r)*} \frac{\mathrm{e}^{\mathrm{i} \int^z \omega_-(z') \mathrm{d}z'}}{\sqrt{2 \omega_-(z)}}, \label{u-}
\end{align}
\end{subequations}
with $\bar r=-r$. We here remind that non-adiabatic frequencies $\omega_\pm$ depend on both the times of oscillation $n$ and the helical state $r$.
Here, $\alpha^{(r)}(k)$ and $\beta^{(\bar{r})}(k)$ are the Bogolyubov coefficients, corresponding to the transmission and reflection amplitudes, respectively, for an incident helical fermion.

Without loss of generality, we will only consider the solution for $u_{r,+}$.
We adopt the convention that the WKB exponent, or the adiabatic phase, takes the expression
\begin{align}
    \Theta^{(r)}(z) = \int_0^z \omega_+(z') \mathrm{d}z', \text{ with } \Theta^{(r)}(z) = -\Theta^{(\bar{r})}(-z) \label{Theta}
\end{align}
The asymmetry relation of $\Theta^{(r)}(z)$ indicates a transition of helical states from the asymptotic past to the asymptotic future.
This is the reason why we use the Bogoliubov coefficients $\beta^{(\bar{r})}$ in Eq. \eqref{uB} instead of $\beta^{(r)}$.
It is clear that, when $|z|\to\infty$, $\Theta^{(r)}$ is dominated by the part $p^2 + (1+\xi^2)z^2$ and the residual part $1/(z \pm  \mathrm{i} r p/\xi)^2$ converses to 0.
Based on these facts, decompose $\Theta^{(r)}$ into two parts $\Theta^{(r)}(z) = \Theta_0(z) + \Theta_1^{(r)}(z)$, where
\begin{align}
    \Theta_0(z) = \int_0^z \sqrt{p^2 + (1+\xi^2)z^2}\, \mathrm{d}z,  \label{Theta_0}
\end{align}
with asymmetry relation $\Theta_0(z) = -\Theta_0(-z)$ being independent of helical states.
The first-order complex phase, denoted as \(\Theta_1^{(r)}(z)\), is the residual part by subtracting $\Theta_0$ from \(\Theta^{(r)}\), which exhibits an antisymmetry as a consequence of relation \eqref{Theta}:
\begin{align}
    \Theta^{(r)}_1(z) = -\Theta^{(\bar{r})}_1(-z).  \label{Theta_1}
\end{align}
In the following sections, we will see that the helical asymmetry arises precisely from the imaginary part of $\Theta^{(r)}_1$.
It is notable that the first-order complex adiabatic phase $\Theta^{(r)}_1(z)$ depends on the times of oscillation $n$ since $\Theta^{(r)}(z)$ and $\Theta_0(z)$ are both dependent on $n$.

Computing the adiabatic phase $\Theta_0(z)$, we obtain the analytical expression
\begin{align}
    \Theta_0 = \frac{p^2}{2 \nu^2} (u + \sinh u \cosh u), \label{Theta0_u}
\end{align}
with $\nu^4 = 1 + \xi^2$ and $z = \frac{p}{\nu^2} \sinh u$. For large $z\to+\infty$, the relevant parameters are asymptotically given by
\begin{subequations}
\begin{align}
    u = \ln \left( \frac{2 \nu^2 z}{p} \right) + \mathcal O(z^{-2}), \\
    \Theta_0 = \frac{\nu^2}{2} z^2 + \frac{p^2}{2 \nu^2} \ln \left( \frac{2 \nu^2 z}{p} \right) + \frac{p^2}{4 \nu^2} + \mathcal O(z^{-2}), \\
    \omega_+ \sim \nu^2 z.
\end{align}
\end{subequations}
Using the antisymmetric relation about $\Theta_0$ in Eq. \eqref{Theta_0}, we have the corresponding form as $z\to-\infty$,
\begin{subequations}
\begin{align}
    \Theta_0(z)\sim -\frac{\nu^2}{2}{z^2}-\frac{p^2}{2\nu^2} \ln \left( \frac{2\nu^2|z|}{p} \right)-\frac{p^2}{4\nu^2}, \\
    \omega_+ \sim \nu^2 |z|.
\end{align}
\end{subequations}
For brevity, we introduce the positive frequency mode $f_r(z)$ and the negative frequency mode $g_r(z)$, which correspond, respectively, to the adiabatic forms in Eq. \eqref{u+}
\begin{align}
    f_r(z) = \frac{\mathrm{e}^{ \mathrm{i} \int^z \omega_+(z') \mathrm{d}z'}}{\sqrt{2\omega_+(z)}}, \
    g_r(z) = \frac{\mathrm{e}^{-\mathrm{i} \int^z \omega_+(z') \mathrm{d}z'}}{\sqrt{2\omega_+(z)}}.
\end{align}
Therefore, the WKB-approximate positive frequency modes for the in-vacuum and out-vacuum states are given by
\begin{subequations}\label{asymptotic_f}
\begin{align}
    &f_r^{\left( \text{in} \right)}\equiv f_r(z\to -\infty) \nonumber\\
    &\quad \sim \frac{1}{\nu \sqrt{2|z|}}\left[\frac{2{\nu }^{2}|z|}{p}\right]^{\frac{\mathrm{i}p^{2}}{2{\nu }^{2}}}
    \mathrm{e}^{\mathrm{i}\frac{{\nu }^{2}}{2}z^{2}}
    \mathrm{e}^{\mathrm{i}\frac{p^{2}}{4{\nu }^{2}}}
    \mathrm{e}^{\mathrm{i}\Theta_{1}^{(r)}(-\infty)}, \\
    &f_{r}^{\left( \text{out} \right)}\equiv f_{r}\bigl(z\to +\infty\bigr) \nonumber\\
    &\ \sim \frac{1}{\nu\sqrt{2z}}\left[\frac{2{\nu }^{2}z}{p}\right]^{-\frac{\mathrm{i}p^{2}}{2{\nu }^{2}}}
    \mathrm{e}^{-\mathrm{i}\frac{{\nu }^{2}}{2}z^{2}}
    \mathrm{e}^{-\mathrm{i}\frac{p^{2}}{4{\nu }^{2}}}
    \mathrm{e}^{\mathrm{i}\,\Theta_{1}^{(r)}(+\infty)^*},
\end{align}
\end{subequations}
where the complex conjugate $*$ on the right hand side of $f_{r}^{(\text{out})}$ comes from the antisymmetric relation \eqref{Theta_1}.

On the other hand, the limiting form of the evolution equation \eqref{evol} can also be written as
\begin{align}
    \frac{\mathrm{d}^2 \tilde{u}_{r,\pm}}{\mathrm{d}z^2}+\left[ p^2\pm \mathrm{i}+ (1+\xi^2)z^2 \right] \tilde{u}_{r,\pm }=0. \label{evol2}
\end{align}
This equation is associated with parabolic cylinder functions $D_{\kappa}(z)$, with the solution given by
\begin{align} \label{f_D}
    f_{r}=\alpha D_{-\frac{1}{2}-\mu}( \sqrt{2}{{\mathrm{e}}^{\mathrm{i}\frac{3}{4}\pi }}\nu z )
          +\beta D_{-\frac{1}{2}+\mu}( \sqrt{2}{{\mathrm{e}}^{\mathrm{i}\frac{5}{4}\pi }}\nu z ),
\end{align}
where the parameter in the order is
\begin{align} \label{mu}
    \mu=\frac{1-\mathrm{i}p^2}{2\nu^2}
\end{align}
with $\nu^4=1+\xi^2$ as defined in Eq. \eqref{Theta0_u}.
To determine the transition of Bogolyubov coefficients from the in-vacuum to the out-vacuum, it is also necessary to analyze the asymptotic behavior of the parabolic cylinder functions.
As $|z| \to \infty$, the asymptotic expansions of the parabolic cylinder function are shown as \cite{NIST}
%\begin{widetext}
\begin{widetext}
\begin{subequations}
\begin{align}
    &D_{\kappa}(z)\sim \mathrm{e}^{-z^2/4}z^{\kappa}-\frac{\sqrt{2\pi}}{\Gamma(-\kappa)}\mathrm{e}^{ \kappa\pi \mathrm{i}}
        \mathrm{e}^{z^2/4} z^{-\kappa-1},\quad \ \ \frac{\pi}{4}<\arg z<\frac{5\pi}{4}, \\
    &D_{\kappa}(z)\sim \mathrm{e}^{-z^2/4}z^{\kappa}-\frac{\sqrt{2\pi}}{\Gamma(-\kappa)}\mathrm{e}^{-\kappa\pi \mathrm{i}}
        \mathrm{e}^{z^2/4} z^{-\kappa -1},\quad \frac{3}{4}\pi <\arg z<\frac{7}{4}\pi.  \label{asymptotic_D}
\end{align}
\end{subequations}
Inserting these asymptotic expressions into Eq. \eqref{f_D} and comparing them with the corresponding asymptotic expansions of $f_r$ in Eq. \eqref{asymptotic_f},
we find that the normalized positive-frequency mode functions associated with the in-vacuum and out-vacuum states are
\begin{subequations}
\begin{align}
    f_{r}^{(\text{in})} &= \frac{1}{\sqrt{\sqrt{2}\nu}} \left(\sqrt{2}{\nu }|z|\right)^{\frac{1}{2\nu^2}} \mathrm{e}^{\mathrm{i}\Theta_1^{(r)}(-\infty)}
        \left(\frac{\sqrt{2}{\nu }}{p}\right)^{\frac{\mathrm{i}p^2}{2\nu^2}} \mathrm{e}^{\mathrm{i}\frac{p^2}{4\nu^2}}\, \mathrm{e}^{-\mathrm{i}\frac{\pi}{4}\left(\frac{1}{2}+\frac{1}{2\nu^{2}}\right)}\,
        \mathrm{e}^{-\frac{\pi}{8}\frac{p^2}{\nu^2}} D_{-\frac{1}{2}-\mu}\left(\sqrt{2}\,\mathrm{e}^{\mathrm{i}\frac{3\pi}{4}}\nu z\right), \\
    f_{r}^{(\text{out})} &= \frac{1}{\sqrt{\sqrt{2}\nu}} \left(\sqrt{2}\nu z\right)^{\frac{1}{2\nu^2}} \mathrm{e}^{\mathrm{i}\Theta_1^{(r)}\left(+\infty\right)^*}
        \left(\frac{\sqrt{2}\nu}{p}\right)^{-\frac{\mathrm{i}p^2}{2\nu^2}} \mathrm{e}^{-\mathrm{i}\frac{p^2}{4\nu^2}}\, \mathrm{e}^{\mathrm{i}\frac{3\pi}{4}\left(-\frac{1}{2}+\frac{1}{2\nu^2} \right)}\,
        \mathrm{e}^{-\frac{\pi}{8}\frac{p^2}{\nu^2}} D_{-\frac{1}{2}+\mu}\left(\sqrt{2}\mathrm{e}^{\mathrm{i}\frac{5}{4}\pi}\nu z \right).  \label{f_out}
\end{align}
By the same method, the normalized negative-frequency mode functions about the in-vacuum and the out-vacuum states are expressed as
\begin{align}
    g_r^{(\text{in})} &= \frac{1}{\sqrt{\sqrt{2}\nu}} \left(\sqrt{2}\nu |z|\right)^{\frac{1}{2\nu^2}} \mathrm{e}^{-\mathrm{i}\Theta_1^{(r)}(-\infty)^*}
        \left(\frac{\sqrt{2}\nu}{p}\right)^{-\frac{\mathrm{i}p^2}{2\nu^2}} \mathrm{e}^{-\mathrm{i} \frac{p^2}{4\nu^2}}\, \mathrm{e}^{- \mathrm{i} \frac{5\pi}{4} \left(-\frac{1}{2} + \frac{1}{2\nu^2}\right)}\,
        \mathrm{e}^{-\frac{\pi}{8} \frac{p^2}{\nu^2}} D_{-\frac{1}{2} + \mu} \left( \sqrt{2} \mathrm{e}^{\mathrm{i} \frac{3}{4} \pi} \nu z \right), \\
    g_r^{(\text{out})} &= \frac{1}{\sqrt{\sqrt{2} \nu}} \left(\sqrt{2}\nu z\right)^{\frac{1}{2\nu^2}} \mathrm{e}^{-\mathrm{i} \Theta_1^{(r)}(+\infty)}
        \left( \frac{\sqrt{2} \nu}{p} \right)^{\frac{\mathrm{i} p^2}{2\nu^2}} \mathrm{e}^{\mathrm{i} \frac{p^2}{4\nu^2}}\, \mathrm{e}^{\mathrm{i} \frac{3\pi}{4} \left( \frac{1}{2} + \frac{1}{2\nu^2} \right)}\,
        \mathrm{e}^{-\frac{\pi}{8} \frac{p^2}{\nu^2}} D_{-\frac{1}{2} - \mu} \left( \sqrt{2} \mathrm{e}^{\mathrm{i}\frac{5}{4} \pi} \nu z \right).   \label{g_out}
\end{align}
\end{subequations}

Recalling the helical-asymmetric dynamics equation \eqref{evol}, the solution for the in-vacuum state, expressed as a combination that matches the asymptotic solution \eqref{f_D} as $z \to -\infty$, is given by
\begin{align} \label{u_in}
    &\tilde{u}_{r,+}^{(\text{in})} = \alpha^{(r)} f_{r}^{(\text{in})} + \beta^{(\bar{r})} g_{r}^{(\text{in})} \nonumber\\
    &=\alpha^{(r)}\frac{1}{\sqrt{\sqrt{2}\nu}} \left(\sqrt{2}{\nu }|z|\right)^{\frac{1}{2\nu^2}} \mathrm{e}^{\mathrm{i}\Theta_1^{(r)}(-\infty)}
        \left(\frac{\sqrt{2}{\nu }}{p}\right)^{\frac{\mathrm{i}p^2}{2\nu^2}} \mathrm{e}^{\mathrm{i}\frac{p^2}{4\nu^2}}\, \mathrm{e}^{-\mathrm{i}\frac{\pi}{4}\left(\frac{1}{2}+\frac{1}{2\nu^{2}}\right)}\,
        \mathrm{e}^{-\frac{\pi}{8}\frac{p^2}{\nu^2}} D_{-\frac{1}{2}-\mu}\left(-(1-\mathrm i)\nu z\right)  \nonumber\\
    &\quad+\beta^{(\bar{r})}\frac{1}{\sqrt{\sqrt{2}\nu}} \left(\sqrt{2}\nu |z|\right)^{\frac{1}{2\nu^2}} \mathrm{e}^{-\mathrm{i}\Theta_1^{(r)}(-\infty)^*}
        \left(\frac{\sqrt{2}\nu}{p}\right)^{-\frac{\mathrm{i}p^2}{2\nu^2}} \mathrm{e}^{-\mathrm{i} \frac{p^2}{4\nu^2}}\, \mathrm{e}^{- \mathrm{i} \frac{5\pi}{4} \left(-\frac{1}{2} + \frac{1}{2\nu^2}\right)}\,
        \mathrm{e}^{-\frac{\pi}{8} \frac{p^2}{\nu^2}} D_{-\frac{1}{2} + \mu} \left(-(1+\mathrm i) \nu z \right).
\end{align}
In the above expression, $\alpha^{(r)}$ and $\beta^{(\bar{r})}$ denote the Bogolyubov coefficients before $\phi$ crosses $\phi_*=0$,
while the functions which multiply them are exact solutions of the linearized equation \eqref{evol2}.
%The analytical approximation also involves treating them as the solutions to the exact evolution equation \eqref{evol}.
To obtain the mode associated with the out-vacuum state, we need the recurrence relations of the parabolic cylinder functions \cite{integrals}
\begin{subequations}
\begin{align}
    D_\kappa(z)&=\mathrm e^{\kappa\pi i}D_\kappa(-z)+\frac{\sqrt{2\pi}}{\Gamma(-\kappa)}\mathrm e^{\pi(\kappa+1)\mathrm i/2}D_{-\kappa-1}(-\mathrm i z) \label{DR_1} \\
    &=\mathrm e^{-\kappa\pi \mathrm i}D_\kappa(-z)+\frac{\sqrt{2\pi}}{\Gamma(-\kappa)}\mathrm e^{-\pi(\kappa+1)\mathrm i/2}D_{-\kappa-1}(\mathrm i z). \label{DR_2}
\end{align}
\end{subequations}
Then, substituting relation \eqref{DR_1} into $D_{-\frac{1}{2}-\mu}\left(-(1 - \mathrm{i}) \nu z\right)$ and
substituting relation \eqref{DR_2} into $D_{-\frac{1}{2}+\mu}\left(-(1 + \mathrm{i}) \nu z\right)$ in the equation above,
respectively, as $z\to+\infty$, it subsequently rewrites the solution \eqref{u_in} in terms of two different parabolic cylinder functions:
\begin{align} \label{u_out}
    \tilde{u}_{r,+}^{(\text{out})}
    =& \left[ \alpha^{(r)} \frac{\sqrt{2\pi}}{\Gamma\left(\frac{1}{2} + \frac{1}{2\nu^2} - \mathrm{i} \frac{p^2}{2\nu^2}\right)}
        \mathrm{e}^{-\frac{\pi}{4} \frac{p^2}{\nu^2}} \left( \frac{\sqrt{2} p}{\nu} \right)^{ {\mathrm{i}p^2}/{\nu^2}}
        \mathrm{e}^{-\mathrm{i} \frac{p^2}{2\nu^2}}\, \mathrm{e}^{-\mathrm{i}\pi\left(\frac12+\frac{1}{2\nu^2}\right) }\,
        \mathrm{e}^{\mathrm{i}[\Theta_1^{(r)}(-\infty) - \Theta_1^{(r)}(+\infty)^*]} \right.     \nonumber\\
    &\quad\quad\quad\quad\quad\quad\quad\quad\quad \left. +\beta^{(\bar{r})} \mathrm{e}^{-\frac{\pi}{2} \frac{p^2}{\nu^2}}\,
        \mathrm{e}^{-\mathrm{i} \pi \left( -\frac{1}{2} + \frac{1}{2\nu^2} \right)}\,
        \mathrm{e}^{-\mathrm{i}[\Theta_1^{(r)}(-\infty)^* + \Theta_1^{(r)}(+\infty)^*]}\right] \times     \nonumber\\
    &\quad\quad \left\{\frac{1}{\sqrt{\sqrt{2}\nu}} \left(\sqrt{2}\nu z\right)^{\frac{1}{2\nu^2}} \mathrm{e}^{\mathrm{i}\Theta_1^{(r)}\left(+\infty\right)^*}
        \left(\frac{\sqrt{2}\nu}{p}\right)^{-\frac{\mathrm{i}p^2}{2\nu^2}} \mathrm{e}^{-\mathrm{i}\frac{p^2}{4\nu^2}}\, \mathrm{e}^{\mathrm{i}\frac{3\pi}{4}\left(-\frac{1}{2}+\frac{1}{2\nu^2} \right)}\,
        \mathrm{e}^{-\frac{\pi}{8}\frac{p^2}{\nu^2}} D_{-\frac{1}{2}+\mu}\left(\sqrt{2}\mathrm{e}^{\mathrm{i}\frac{3}{4}\pi}\nu z \right) \right\}  \nonumber\\
    & + \Bigg[ -\alpha^{(r)} \mathrm{e}^{-\frac{\pi}{2} \frac{p^2}{\nu^2}}\, \mathrm{e}^{\mathrm{i} \pi \left( -\frac{1}{2} + \frac{1}{2\nu^2} \right)}\,
        \mathrm{e}^{\mathrm{i} [\Theta_1^{(r)} (-\infty) + \Theta_1^{(r)} (+\infty)]}  \nonumber\\
    &\quad\quad\quad\quad \left. + \beta^{(\bar{r})} \frac{\sqrt{2\pi}}{\Gamma \left( \frac{1}{2} + \frac{1}{2\nu^2} +\mathrm{i} \frac{p^2}{2\nu^2} \right)}
        \mathrm{e}^{-\frac{\pi}{4} \frac{p^2}{\nu^2}} \left( \frac{\sqrt{2}\nu}{p} \right)^{-\mathrm{i}  \frac{p^2}{\nu^2}}
        \mathrm{e}^{\mathrm{i}\frac{p^2}{2\nu^2}}\, \mathrm{e}^{\mathrm{i}\pi\left(\frac12+\frac{1}{2\nu^2}\right) }\,
        \mathrm{e}^{-\mathrm{i} [ \Theta_1^{(r)}  (-\infty)^* - \Theta_1^{(r)}  (+\infty)]} \right]\times \nonumber\\
    &\quad\quad \left\{ \frac{1}{\sqrt{\sqrt{2} \nu}} \left(\sqrt{2}\nu z\right)^{\frac{1}{2\nu^2}} \mathrm{e}^{-\mathrm{i} \Theta_1^{(r)}(+\infty)}
        \left( \frac{\sqrt{2} \nu}{p} \right)^{\frac{\mathrm{i} p^2}{2\nu^2}} \mathrm{e}^{\mathrm{i} \frac{p^2}{4\nu^2}}\, \mathrm{e}^{\mathrm{i} \frac{3\pi}{4} \left( \frac{1}{2} + \frac{1}{2\nu^2} \right)}\,
        \mathrm{e}^{-\frac{\pi}{8} \frac{p^2}{\nu^2}} D_{-\frac{1}{2} - \mu} \left( \sqrt{2} \mathrm{e}^{\mathrm{i}\frac{5}{4} \pi} \nu z \right)  \right\} .
\end{align}

In this new expression, the functions enclosed within the curly brackets correspond to the positive and negative modes of the out-vacuum,
denoted by $f_{r}^{(\text{out})}$ and $g_{r}^{(\text{out})}$ respectively, which are expressed in Eqs. \eqref{f_out} and \eqref{g_out}.
The coefficients preceding them thus represent the new Bogolyubov coefficients expressed in terms of the initial ones.
The derivation above concerns a specific production event (labeled by index $n$) that occurs whenever the pseudo-mass $m_5$ crosses zero.
When considering successive productions, this entire derivation can be easily generalized to the successive zeros of the pseudo-mass.
The important points we only consider are:
(1) a difference in the values of the scale factor $a$ and of the derivative $\phi$ at different $\tau_{*n}$'s;
(2) a change of sign about the parameter $\xi_n$ whenever $\phi$ crosses $\phi_*=0$ from below to above.
Then, we have the transition relations for the Bogolyubov coefficients $\alpha_n^{(r)}$ and $\beta_n^{(\bar{r})}$ within a specific production process
\begin{align}  \label{transform}
    \begin{pmatrix}
    \alpha_n^{(r)} \\[1mm]
    \beta_n^{(\bar{r})}
    \end{pmatrix}
    =
    \begin{pmatrix}
    F_n^{(r)} & H_n^{(r)} \\[1mm]
    -H_n^{(r)*} & F_n^{(r)*}
    \end{pmatrix}
    \begin{pmatrix}
    \alpha_{n-1}^{(r)} \\[1mm]
    \beta_{n-1}^{(\bar{r})}
    \end{pmatrix}
    \text{      for $n$ odd}, \nonumber\\
    \qquad\qquad\qquad\qquad \xi_n=-\xi_n\qquad\qquad\qquad \text{ for $n$ even},
\end{align}
where the coefficients in the transition matrix read
\begin{align}
    &F_n^{(r)} = \frac{\sqrt{2\pi}}{\Gamma\left(\frac{1}{2} + \frac{1}{2\nu_n^2} + \mathrm{i}\frac{p_n^2}{2\nu_n^2}\right)}
    \exp\left\{-\frac{\pi}{4} \frac{p_n^2}{\nu_n^2}+\mathrm{i} \left[ \Theta_1^{(r)}(-\infty) - \Theta_1^{(r)}(+\infty)^* \right] \right\}\cdot \nonumber\\
    &\qquad\qquad\qquad\qquad\qquad\qquad\qquad\qquad \exp\left\{\mathrm i\left[-\frac{p_n^2}{2\nu_n^2}+\frac{p_n^2}{\nu_n^2}\ln\frac{\sqrt{2} p_n}{\nu_n}
        -\pi\left(\frac12+\frac{1}{2\nu_n^2}\right) \right]\right\}, \nonumber\\
    &H_n^{(r)} = \exp\left\{-\frac{\pi}{2} \frac{p_n^2}{\nu_n^2}-\mathrm{i} \left[ \Theta_1^{(r)}(-\infty)^* + \Theta_1^{(r)}(+\infty)^* \right] \right\}
    \exp\left\{-\mathrm{i}\pi \left( -\frac{1}{2} + \frac{1}{2\nu_n^2} \right) \right\}. \label{FG}
\end{align}
We remind that the parameters are defined as $p_n\equiv k/\sqrt{m_\psi a'_{*n}}$ and $\nu_n\equiv(1+\xi_n)^{1/4}$.\\

\section{Helical fermionic density \label{number}}

\subsection{Successive helicity production}
The occupation number of created helical fermions after the $n$-th successive production is given by
\begin{align}
    N_n^{(r)}(k)=\left| \beta _n^{(r)}(k) \right|^2. \label{N}
\end{align}
With the application of the iterative equation for the Bogoliubov coefficients \eqref{transform}, one can get that the $n$-th scattering gives
\begin{align}
    N_{n+1}^{(r)}=&\mathrm{e}^{-\pi \kappa_{r,n}^2}-\mathrm{e}^{-\pi \kappa_{r,n}^2}N_n^{(\bar r)} %\nonumber\\
         +\frac{2\pi}{\left|\Gamma\left(\frac{1}{2} + \frac{1}{2\nu_n^2} + \mathrm{i}\frac{p_n^2}{2\nu_n^2}\right) \right|^2}\exp\left\{-\frac\pi 2 \chi_{r,n}^2\right\} N_n^{(r)} \nonumber\\
         &-2\exp\left\{-\frac\pi 2 \kappa_{r,n}^2-\frac\pi 4 \chi_{r,n}^2\right\}
            \frac{\sqrt{2\pi}}{\left|\Gamma\left(\frac{1}{2} + \frac{1}{2\nu_n^2} + \mathrm{i}\frac{p_n^2}{2\nu_n^2}\right) \right|}\cdot %\nonumber\\
         \sqrt{N_n^{(r)}(1-N_n^{(\bar r)})}\sin\theta_n, \label{N_pm}
\end{align}
\end{widetext}
%\twocolumn
where $\theta_j$ is the total phase accumulated by the moment $t = t_{*j}$.
The parameters in the exponentials are expressed as
\begin{align}
    \kappa _{r,n}^{2}
    &=\frac{p_n^2}{\nu_n^2}+\frac{2}{\pi }\mathrm{Im}\left[ \Theta _1^{(r)}(-\infty)^{*}+\Theta _1^{(r)}( +\infty)^{*} \right] \nonumber\\
    &=\frac{p_n^2}{\sqrt{1+\xi _{n}^{2}}}-\frac{2}{\pi }\mathrm{Im}\left[ \Theta _{1}^{\left( r \right)}\left( -\infty  \right)-\Theta _{1}^{\left( {\bar{r}} \right)}\left( -\infty  \right) \right] \nonumber\\
    &=\frac{p_n^2}{\sqrt{1+\xi _{n}^{2}}}+\frac{2}{\pi }\mathrm{Im}\left[ \Theta _{1}^{\left( {\bar{r}} \right)}\left( +\infty  \right)-\Theta _{1}^{\left( r \right)}\left( +\infty  \right) \right], \label{kappa_r}
\end{align}
and
\begin{align}
    \chi _{r,n}^{2}
    &=\frac{p_n^2}{\nu_n^2} - \frac{4}{\pi }\mathrm{Im}\left[ \Theta_1^{(r)}(-\infty) - \Theta_1^{(r)}(+\infty)^* \right] \nonumber\\
    &=\frac{p_n^2}{\sqrt{1+\xi _{n}^{2}}}-\frac{4}{\pi }\mathrm{Im}\left[ \Theta _{1}^{\left( r \right)}\left( -\infty  \right)-\Theta _{1}^{\left( {\bar{r}} \right)}\left( -\infty  \right) \right].
\end{align}
In the above two equations, we have used the antisymmetric relation \eqref{Theta_1} about $\Theta _{1}^{(r)}\left( z \right)$.
Physically, the first term on the right-hand side of Eq.~\eqref{N_pm} corresponds to the particle distribution in the absence of coherence between oscillations and exhibits a modified Gaussian profile. 
The remaining terms arise from the cumulative effects of previous oscillations and reflect the coherence between successive production events.

If one starts with no fermions production initially, it indicates $\alpha _0^{(\pm)}=1/\sqrt{2}$ and $\beta _0^{(\pm)}=0$.
Then, applying successive transfer matrix \eqref{transform}, we get the spectra of fermions produced after every $t_{*n}$.
Specially, for the case with single oscillation, our calculation reproduces the result directly
\begin{align}
    N_{1}^{(r)} = \frac12\mathrm{e}^{-\pi \kappa_{r,1}^2}.
\end{align}
If we further consider the vanishing of the pseudoscalar term with \( g=0 \) (i.e., \(\nu_n=1\), \(\xi=0\), and \(\Theta^{(r)}_1(\pm\infty)=0\) accordingly), we arrive at a concise result
\begin{align}
    N_{1} = \mathrm{e}^{-\pi k^2/m_\psi a'}, \label{Gaussian}
\end{align}
which have been widely reported in case of helical-symmetric fermion productions \cite{preheating_fermion1}.

\subsection{Semi-analytical result \label{semi}}

The same convention used in the bosonic case \cite{preheating_boson1} also proves useful simplification for the calculations:
when the expansion of the universe is taken into account, the phases in Eq. \eqref{FG} are uncorrelated.
The “random walk recipe” suggests that the best estimate of the above quantity is obtained by summing the squares of the contributions from each oscillation, since the cross terms (mixed products) average to zero. 
This approximation assumes that successive oscillations are uncorrelated, i.e., incoherent with one another. 
Consequently, the amplitude of helical fermions after the $n$-th production event is given solely by the first term on the right-hand side of Eq.~\eqref{N_pm}, while the remaining terms are neglected. 
With this method, Eq. \eqref{FG} is simplified to the following form:
\begin{align} \label{FG2}
    \begin{pmatrix}
    |\alpha_n^{(r)}|^2 \\[1mm]
    |\beta_n^{(\bar{r})}|^2
    \end{pmatrix}
    =
    \begin{pmatrix}
    |F_n^{(r)}|^2 & |H_n^{(r)}|^2 \\[1mm]
    |H_n^{(r)}|^2&  |F_n^{(r)}|^2
    \end{pmatrix}
    \begin{pmatrix}
    |\alpha_{n-1}^{(r)}|^2 \\[1mm]
    |\beta_{n-1}^{(\bar{r})}|^2
    \end{pmatrix}
\end{align}
where the squared norms, briefly, are given by
\begin{align}
    |F_n^{(r)}|^2 = 1 - \mathrm{e}^{-\pi \kappa_{r,n}^2},\quad |H_n^{(r)}|^2 =\mathrm{e}^{-\pi \kappa_{r,n}^2},
\end{align}	
with $|F_n^{(r)}|^2 + |H_n^{(r)}|^2 = 1$. The parameter $\kappa_{r,n}^2$ is already defined in Eq. \eqref{kappa_r}.
Again, we remind that the first-order adiabatic phases $\Theta _1^{(r)}(\pm\infty)$ are dependent on the oscillation time $n$.
It is evident from Eq. \eqref{kappa_r} that the helicity of Dirac fermions during preheating arises entirely from the imaginary part of first-order adiabatic phase $\Theta _1^{(r)}$,
which works in the non-adiabatic region where $z\sim0$.

However, to obtain the analytical result of $\Theta _1^{(r)}(\pm\infty)$ is impossible. One may use the Taylor expansion:
\begin{align}
    &\Theta _{1}^{(r)}(-\infty) \nonumber\\
    &=\int_0^{-\infty}\sqrt{p_n^2+(1+\xi_n^2)z^2+1/4(z+\mathrm{i}rp_n/\xi_n)^2}\mathrm{d}z   \nonumber\\
    &\quad -\int_0^{-\infty}\sqrt{p_n^2+(1+\xi_n^2)z^2}\mathrm{d}z         \nonumber\\  % \label{Theta1_analy}  \\
    &\approx \frac18\int_0^{-\infty}\frac{\mathrm{d}z}{\sqrt{p_n^2+(1+\xi_n^2)z^2}\, (z+\mathrm{i}rp_n/\xi_n)^2}   \nonumber\\
    &= \frac{\mathrm{i}}{8}\frac{r\xi _n^3}{p_n^2}+\text{real part}. \label{Theta1_approx}
\end{align}
In the case of an expanding universe, successive phases can be considered random, analogous to the situation described in Ref. \cite{preheating_fermion1}.
We therefore simply quote the result here.
Assuming no fermions in the initial state, i.e., $\alpha _0^{(\pm)}=1/\sqrt{2}$ and $\beta _0^{(\pm)}=0$,
after $n$ successive production events, the smoothed particle occupation number of $r$-helicity is given by
\begin{align}
    &N_n^{(r)}(k)=\left| \beta _n^{(r)}(k) \right|^2  \nonumber\\
    &=\frac{1}{4}-\frac{1}{4}\prod_{i=1}^n\left( 1-2{{\mathrm{e}}^{-\pi \kappa _{r,n}^{2}}} \right)\nonumber\\
    &\approx \frac{1}{2}\sum\limits_{i=1}^{n}\mathrm{e}^{-\pi \kappa _{r,n}^2}
\end{align}

The helicity of Dirac fermions could be quantitatively characterized by the net helical density, which represents the density of left-handed helical particles minus the right-handed helical particles.
The net density after first $n$'s productions reads
\begin{align}
    &Q_n=\frac{1}{2\pi^2}\int_0^{\infty}\mathrm d k\ k^2 \left[ N_n^{(+)}(k)-N_n^{(-)}(k) \right] \nonumber\\
    &= \sum_{i=1}^{n} \frac{(m_\psi a'_{*j})^{3/2}}{4\pi^2} \sum_{r=\pm} r\int_0^\infty\mathrm d p_j\, p_j^2 \mathrm{e}^{-\pi \kappa_{r,j}^2}.   \label{Qn}
\end{align}
Then the net density generated with infinite oscillations becomes
\begin{align}
    &Q=\lim_{n \to \infty}Q_n  \nonumber\\
    &={\lim\limits_{n \to \infty }}\sum_{j=1}^{n} \frac{( m_\psi a'_{*j})^{3/2}}{4\pi^2}{I_j}. \label{Q1}
\end{align}
Here, $I_j$ contains an integral involving the phase $\Theta _1^{(r)}(\pm\infty)$.
We retain only the imaginary part in Eq. \eqref{Theta1_approx} for brevity in further calculations.
Then the expression of $I_j$ writes
\begin{align}
    I_j &= \sum_{r=\pm}r \int_0^\infty \mathrm{d}p \, p^2 \exp\left(-\frac{\pi p^2 }{\sqrt{1+\xi_j^2}}+\frac{1}{2\pi} \frac{r \xi_j^3}{p^2}\right) \nonumber\\
    &=\sum_{r=\pm}\int_0^\infty \mathrm{d}p \, p^2\mathrm{e}^{-\pi p^2 / \sqrt{1+\xi_j^2}} \sum_{m=0}^\infty \frac{1}{m!} \left( \frac{1}{2\pi} \frac{r \xi_j^2}{p^2} \right)^m \nonumber\\
    &= \pi^{-1} \xi_j^3 \int_0^\infty \mathrm{d}p \, \mathrm{e}^{-\pi p^2 / \sqrt{1+\xi_j^2}} + \mathcal O(\xi_j^6) \nonumber\\
    &= \frac{1}{2\pi} \xi_j^3 (1+\xi_j^2)^{1/4}.  \label{Ij}
\end{align}
The divergent terms in the third equality, which arise from the lower limit of integration, have been neglected. 
The precise upper integration limit is the helicity-dependent chemical potential $\mu_r$ \cite{axion1},
which, in local thermal equilibrium, arises from the pseudoscalar coupling between the fermion fields and the oscillating inflaton;
nevertheless, when $\mu_r$ is much larger than the static fermion mass $m_\psi$, the approximation \eqref{Ij} remains well justified.    

In this work, preheating is considered as the early stage of the radiation-dominated era, quantized by the proportional relation $a \propto \tau$ or $a\propto t^{1/2}$.
Together with the expression of the oscillating inflaton \eqref{phi}, the dimensionless parameters then become
\begin{align}
    p_j=\text{Constant},\quad \xi_j = \frac{g M_p}{\sqrt{3\pi^3} f} \cdot \frac{(-1)^{j-1}}{j}, \label{p_xi}
\end{align}
where we have set $t_{*j}=\pi j/m_\phi$ whenever the helicity production events occur.
Inserting Eqs. \eqref{Ij} and \eqref{p_xi} into Eq. \eqref{Q1}, we finally have the quasi-analytical expression about the final helical density
\begin{align}
    &Q =\sum_{j=1}^{\infty}(-1)^{j-1}\frac{ (m_\phi a'_{*})^{3/2}}{(2\pi)^3} \left( \frac{\xi_1}{j} \right)^3
    \left[ 1 + \left(\frac{\xi_1}{j} \right)^2 \right]^{\frac14}  \nonumber\\
    &\approx \frac{ (m_\phi a'_{*})^{3/2}}{(2\pi)^3} \cdot \begin{cases}\xi_1^3\eta(3),&\xi_1\ll1,\\[1mm] \xi_1^{7/2}\eta\left(\frac{7}{2}\right),&\xi_1\gg1,\end{cases} \label{Q2}
\end{align}
with $\xi_1=g M_p/\sqrt{3\pi^3}f$ and $\eta(n)=\sum_{j=1}^\infty (-1)^{j-1} /j^n $.

We emphasize here that the quasi-analytical results in this section are based on the assumption of the "random walk recipe" where the coherence between oscillations is neglected.
However, this approximation provides a directly  description of the helical asymmetry. For more accurate results regarding helical production, a detailed analysis will be carried out in the next section using numerical simulations. 

\section{Numerical results \label{Num}}

In this section, we focus on characterizing the occupation number of helical fermions produced during the preheating phase.
Preheating is assumed to occur in the early radiation-dominated phase, so that the parameters $p_j$ and $\nu_j$ obtained in Sec. \ref{semi} can be directly applied.
Helical fermion production occurs throughout inflation and the subsequent epochs; however, a significant net helicity imbalance emerges only during the preheating stage.
Contrary to the conventional expectation that inflation itself generates this imbalance, we show that the asymmetry is instead created by the non-adiabatic dynamics of preheating,
and its final magnitude is determined within the first 50 inflaton oscillations.

\subsection{Static universe}

\begin{figure}
	\centering
	\includegraphics[width=\linewidth]{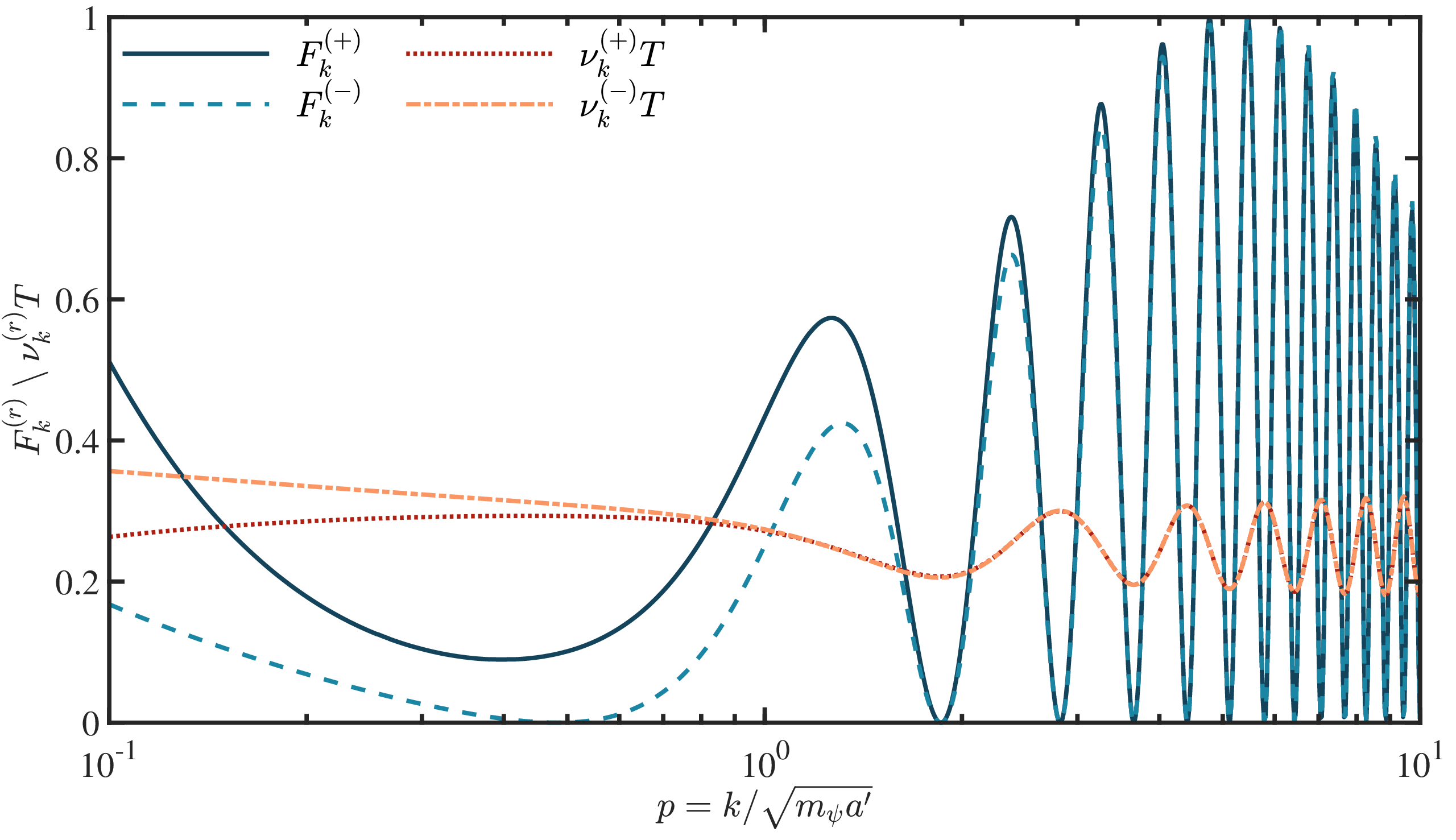}
	\caption{The fermion spectrum $F^{(r)}_k$ and $\nu^{(r)}_k T$ as a function of $p$, with $\xi_1=1.25$.  }
	\label{F_r}
\end{figure}
We first consider a simple scenario in which the universe remains in a static state.
In this approximation, we treat the parameters $p_n$ and $\xi_n$ as constants and rewrite them as $p$ and $\xi_1$ for brevity.
The simplicity of the model enables a clean understanding on the successive production of asymmetric fermions.

Following the analysis in Refs. \cite{preheating_fermion5,preheating_fermion3,static1},
the square of the coefficient of the negative frequency mode represents the particle number and can be extracted through
\begin{equation}
    \bar{N}^{(r)}_k(t)=\frac{1}{T}\int_t^{t+T}N^{(r)}_k(t)\mathrm{d}t=F_k^{(r)}\sin^2(\nu_kt),
\end{equation}
where $T$ is the oscillating period of $\phi(t)$. The amplitude $F^{(r)}_k$ and frequency $\nu_k$ are given by
\begin{subequations}
\begin{align}
    F^{(r)}_k=\frac{1}{\sin^2(\nu^{(r)}_k T)}\frac{p^2}{|\omega|^2}\left[\mathrm{Im}\left(Z^{(r)}_k(T)\right)\right]^2, \\
    \cos(\nu^{(r)}_k T)=-\mathrm{Re}\left(Z^{(r)}_k(T)\right).  \label{F_nu}
\end{align}
\end{subequations}
Here the complex frequency is defined as
\begin{align}
    \omega^{2}(z;r) = p^2 +\mathrm{i} + (1+\xi^2) z^{2} +\frac{1}{4(z+\mathrm{i}rp/\xi)^{2}}. \label{omega_r}
\end{align}
The function $Z^{(r)}_k(t)$ is the first fundamental solution of the equation of motion \eqref{evol}, meaning that $Z^{(r)}_k|_{z=0}=1$ and $\partial_z Z^{(r)}_k|_{z=0}=0$.
Note, however, that the occupation number $N^{(r)}_k$ is only valid if the WKB approximation holds, i.e., $\partial_z\omega/\omega^2\ll1$.

\begin{figure}
	\centering
	\includegraphics[width=\linewidth]{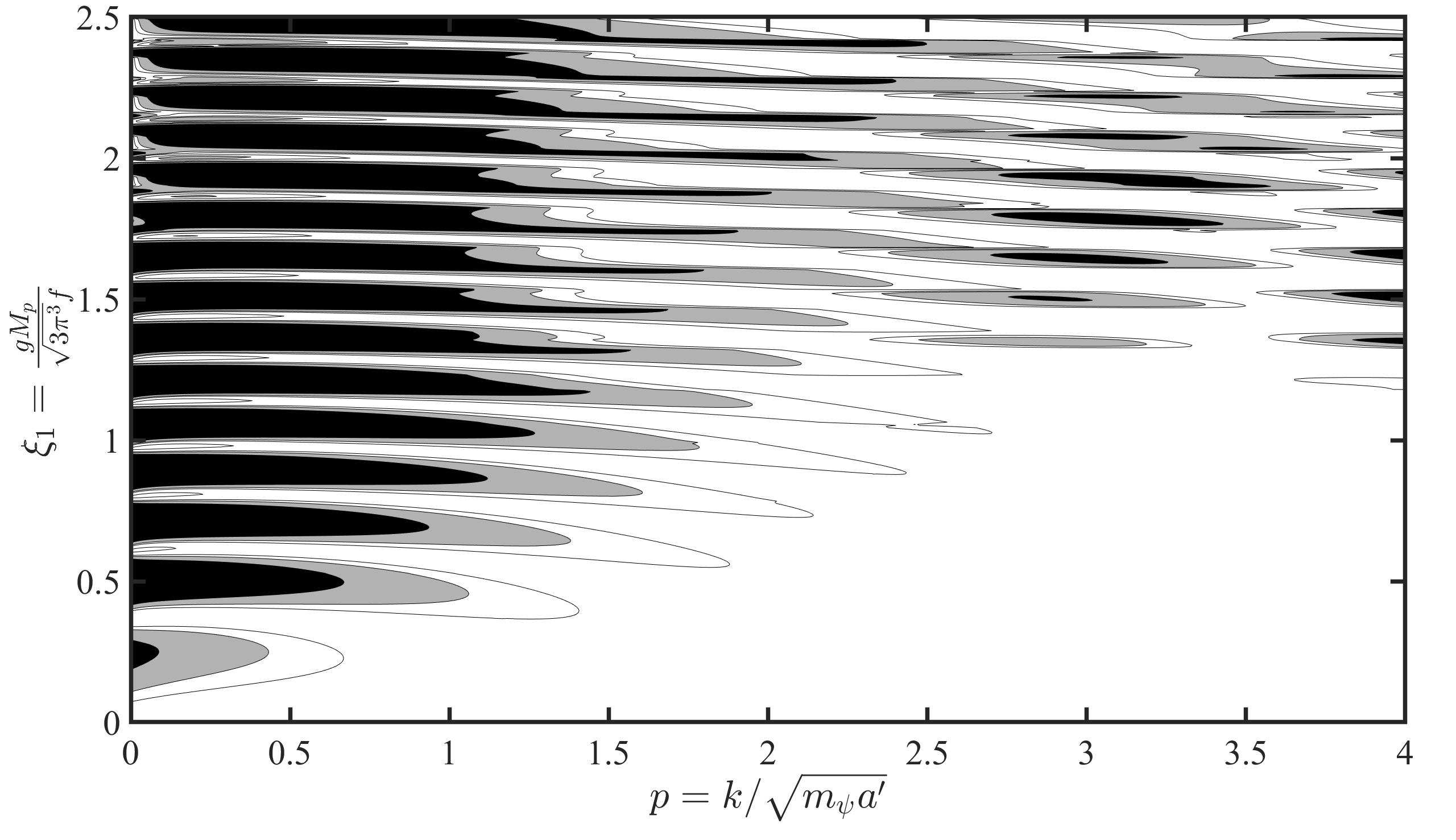}
	\caption{The instability chart for helicity. Contours correspond to equipotential values of $\Delta F_k=F^{(+)}_k-F^{(-)}_k$=0.05, 0.1, and 0.15 in the plane $(p,\xi_1)$ from lighter to darker.  }
	\label{Delta_F}
\end{figure}
\begin{figure*}
	\centering
	\includegraphics[width=\linewidth]{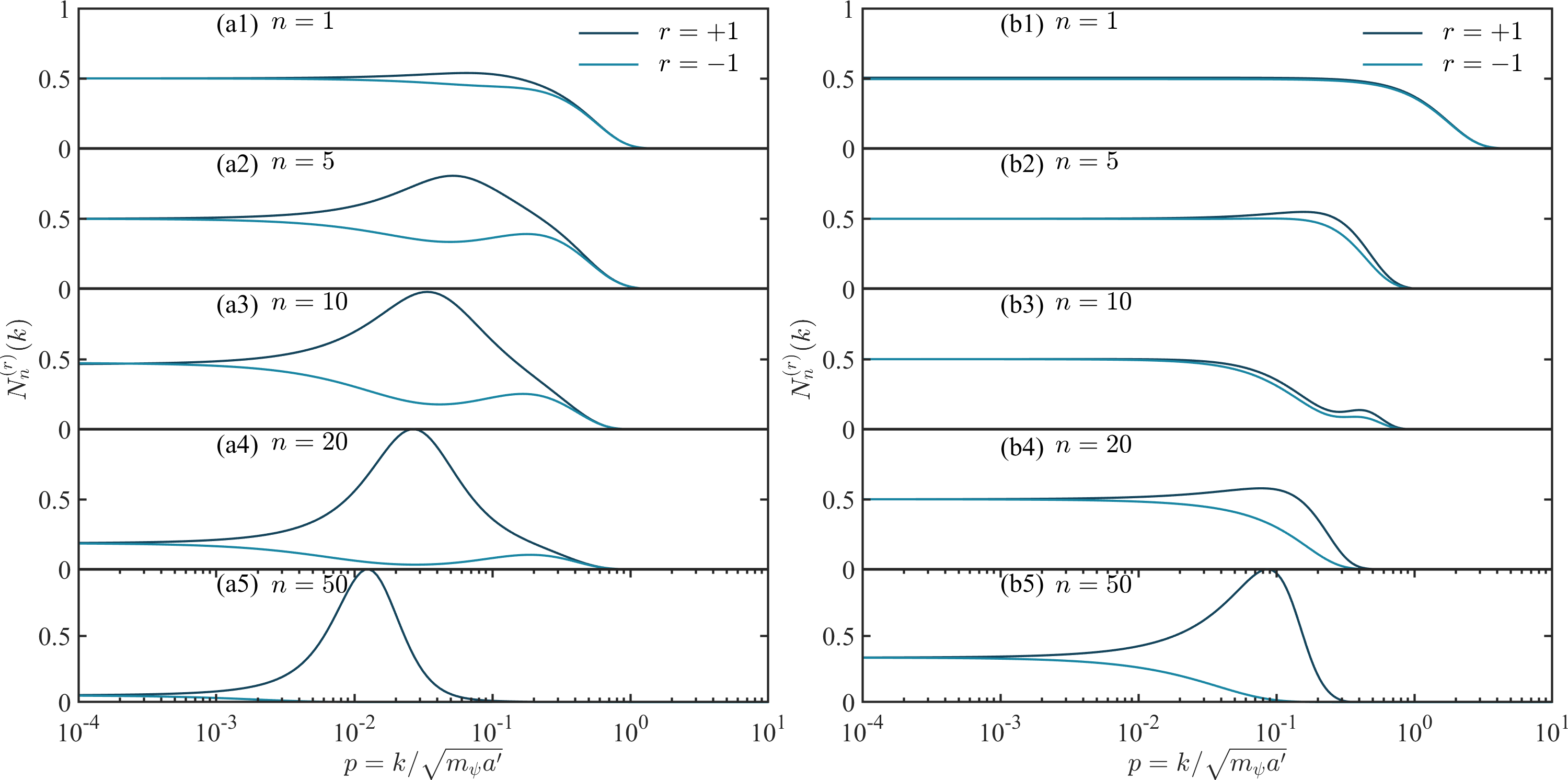}
	\caption{Occupation number density distributions $N_n^{(r)}(k)$ of each helicity for (left) $\xi_1=0.1$ and (right) $\xi_1=1$.
            The panels correspond to the times of oscillation $n=1$, 5, 10, 20, and 50 from top to bottom, respectively. }
	\label{expansion_N_n_k}
\end{figure*}

Figure \ref{F_r} shows the behaviors of $F_k^{(r)}$ and $\nu_k^{(r)}$ for $\xi_1=1.25$.
It clearly demonstrates that the amplitudes of different helicities $F_k^{(r)}$ vary from one another, particularly in the region where $p \sim 1$.
However, they exhibit a decreasing trend of their difference as $p$ increases.
This indicates that the femionic helical asymmetry is primarily influenced at large scale.
In logarithmic momentum space, the shape of $F_k^{(r)}$ also exhibits several narrow peaks.
This suggests that the generation of helicity would occur solely through resonance bands.
In addition, the peaks of $F_k^{(+)}$ and $F_k^{(+)}$ always locate at the same positions, corresponding to the instability for a fixed value of $\xi_1$.
To describe the asymmetry directly, we show in Fig. \ref{Delta_F} the instability chart for the difference of amplitude $\Delta F_k=F^{(+)}_k-F^{(-)}_k$.
It displays the contours of equal $\Delta F_k$ values in the $(p, \xi_1)$ plane.
Fermionic helical asymmetry primarily arises from momenta located in the darker regions, which appear as isolated narrow bands near the maxima of $F_k^{(r)}$.
After several bands, these bands shrink to a negligible area as $p$ increases.

\subsection{Expanding universe}

The static universe approximation provides a more intuitive explanation for the fermionic asymmetry generated by pseudoscalar preheating.
However, beyond the individual particle production events, it does not accurately illustrate the phenomenon from such an epoch.
In this section, the numerical results are applied to the exact analytical expressions given in Eq. \eqref{transform}.

As demonstrated in Eq. \eqref{Gaussian}, when the pseudoscalar coupling vanishes ($g=0$), the occupation numbers in momentum space $k$ exhibit identical Gaussian distributions for each helicity.
The inclusion of the pseudoscalar term, however, alters these distributions.
Figure \ref{expansion_N_n_k} illustrates that, for both positive and negative helical modes, the fermionic occupation number distribution \eqref{N} exhibits still a damping trend.
Specifically, this trend manifests as oscillatory attenuation, which can be verified in a logarithmic scale on the $y$-axis.
Clearly, this deviation for each helicity arises from the imaginary part of the first-order complex adiabatic phase $\Theta^{(r)}_{1}$, as expressed in Eq. \eqref{Theta1_approx}.
The left panels (a1)-(a5) illustrate the distributions for the case where $\xi_1=0.1$ after oscillations with $n=1$, 5, 10, 15, and 40 from top to bottom.
It is evident that as the number of successive production events increases, the distributions increasingly deviate from a Gaussian format and
the curves seperate from each other more significantly.
It is also interesting to see that this difference only appears at a specific wavelength which corresponds to the distribution peak.
In addition, as productions increase, the peaks move to a smaller wavenumber $p$ which locates at $8\xi_n$ approximately.

Figure \ref{expansion_N_n_k}(b1)-(b5) plots the occupation number distributions for both positive and negative helicities, with the parameter set to $\xi_1=1$.
We see that, although the parameters $\xi_1$ differ, similar conclusions can be drawn by comparing those in Fig. \ref{expansion_N_n_k}(a1)-(a5).

Figures \ref{expansion_N_n_k} in both left and right panels implies the
the first few production events are important for determining the characteristics of the particle spectrum at small scale $p>1$.
Subsequent production events can only affect on the smaller wavenumbers with a increasing strength.
Based on the above conclusions, we infer that when $\xi_1\ll1$, the occupation number distributions are mainly mainly concentrated at super-horizon scales $p\ll1$.
This results in a vanishing radial dependence of the distribution, $k^{2}N_{n}^{(r)}(k)\to0$, yielding almost equal comoving number densities for the two helicity states.

For bosons, occupation numbers grow exponentially due to Bose enhancement and parametric resonance~\cite{preheating_boson2}. 
In contrast, fermions obey Fermi-Dirac statistics, which restricts the occupation number of each mode to be at most unity. 
In the fermionic case, the combination of parametric resonance and Pauli blocking leads to coherent enhancement in certain frequency bands and coherent suppression in others, as illustrated in Fig.~\ref{expansion_N_n_k}.
As the number of oscillations increases, the difference between the two helicity states grows prominently around $p \approx 8\xi_n$. 
Meanwhile, in other regions, particularly near $p \sim 0$ and $p \gtrsim 1$, the distributions for both helicities tend to coincide and exhibit nearly Gaussian profiles. 
While Pauli exclusion prevents the exponential growth seen in the bosonic case, coherence with previous production events modifies the present distribution, resulting in enhancement or compression within specific momentum ranges.
This behavior is also evident in the static universe, where parametric resonance of the helical asymmetry manifests as isolated narrow bands. 
To further clarify these coherence effects, we analyze the time evolution of the number densities for each helicity state below.

\begin{figure*}
	\centering
	\includegraphics[width=\linewidth]{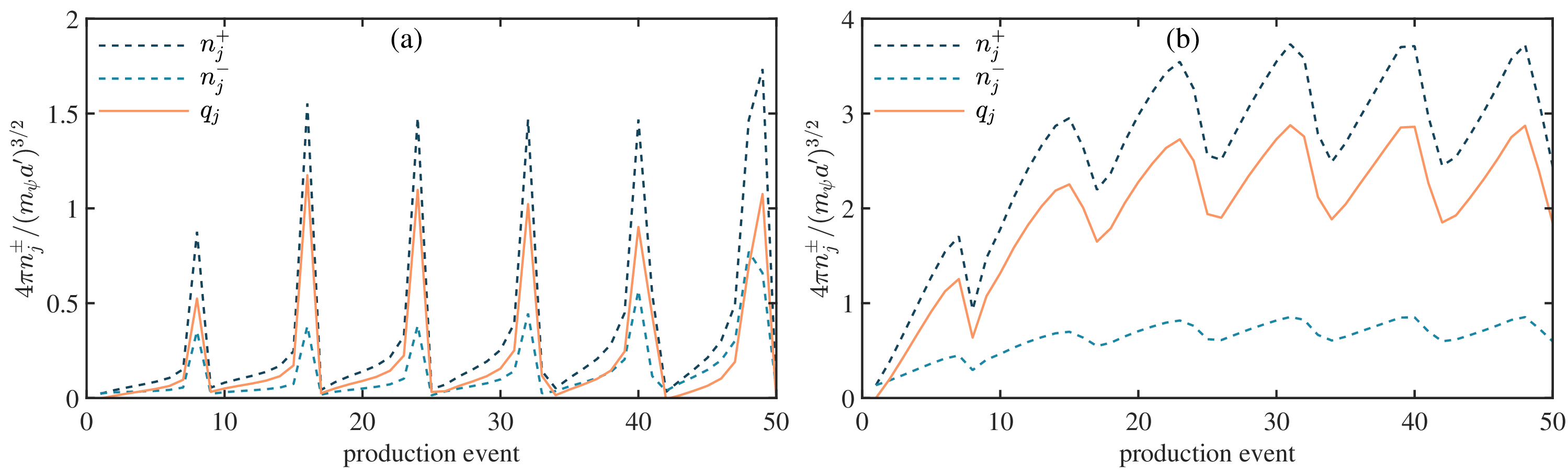}
	\caption{ Evolutions of comoving number densities for different helicities after successive production events with (a) $\xi_1=0.1$ and (b) $\xi_1=1$.
            The net density is defined as $q_j\equiv n_j^+-n_j^-$.}
	\label{occupation}
\end{figure*}

We next examine the time evolution of the comoving number densities for each helicity state $r$, defined as
\begin{align}
    n_j^\pm= \frac{1}{2\pi^2}\int\mathrm dk\,k^2N_n^{(\pm)}(k).  \label{n}
\end{align}
The numerical results are illustrated in Fig. \ref{occupation}, in which the left panel is the number density in the case of $\xi_1=0.1$ while the right corresponds to $\xi_1=1$.
It shows that the evolutions of net density $q_j\equiv n_j^+-n_j^-$ manifest as an oscillating trend, with an approximate periodicity of 8.
This phenomenon may arise from the factor $1/8$ in the first-order complex adiabatic phase $\Theta^{(r)}_{1}$, as expressed in Eq. \eqref{Theta1_approx}.
The real part yields an imaginary phase by multiplying with the unit imaginary number, thereby enhancing the productions through coherent superposition after eight oscillations.
For smaller values of $\xi_1$, which indicates a weaker coupling constant $g$, there is a dramatic production of helical asymmetry after a certain number of oscillations, but the amplitude is significantly suppressed.
On the contrary, for larger values of $\xi_1$, the successive productions become more gradual, which also predicts a significant amplitude of the net density $q_j$.
Physically, the oscillation of number densities comes from the competition between the parametric resonance and the Pauli blocking, 
which differs dramatically from the exponential growth of the bosen case. 

In the context of an expanding universe, the oscillatory behavior of the average particle number is suppressed.
This suppression occurs because particle production following each event is less efficient than that of the preceding event.
This inefficiency arises from the fact that the amplitude and consequently the velocity of the inflaton oscillations is dampened by Hubble friction.
Therefore, we would like to emphasize that Fig. \ref{occupation} illustrates the evolution of net helical number densities in comoving coordinates.
The physical densities are obtained by multiplying by the factor $a_{*j}^{-3}$.

Finally, we would like to discuss the effects from the static fermion mass $m_\psi$.
The massless fermions ($m_\psi=0$) makes the disappearance of the term $\pm i$ in Eqs. \eqref{evol}, which differs significantly from the bosonic case.
In this work, a similar conclusion would also be drawn for massless fermions.
As shown in Refs. \cite{fermion2,massless}, there is no gravitational particle production for massless fermions in an exact Friedmann-Robertson-Walker geometry.
This result can also be verified using Eq. \eqref{Qn}.
Occupation numbers in Fig. \ref{expansion_N_n_k} are scaled by the characteristic wavenumber $k_*^2=m_\psi a'$, which means fermions with heavier mass prefer to distribute on a smaller physical scale.

\section{Conclusions and Discussions \label{conclusion}}

The successive production of fermions without helical asymmetry during preheating has been demonstrated previously, showing a strong similarity to the preheating of bosons.
However, discussions regarding the generation of fermionic helical asymmetry, particularly the analytical results, have not been thoroughly conducted.
In this study, we investigate the asymmetry during preheating by introducing a model in which the Dirac field couples to the scalar inflaton through a pseudoscalar mechanism.
Following inflation, as the inflaton oscillates, the effective frequency of the fermion field varies non-adiabatically, resulting in particle production.
These oscillations also indicate that the pseudo-mass $m_5$, as defined in Eq. \eqref{m5}, changes sign, resulting in the production of unequal fermion helicities.
With the applications of WKB method, the transition relations of the Bogoliubov coefficients are calculated analytically, which involves the parabolic cylinder functions.
Unlike the scenario without the pseudoscalar term, an additional complex phase arises, which directly influences fermion helicity.
The formulae we obtained is valid for an arbitrary number of productions. This is the primary focus of the first part of the present work.

The numerical results show that the production of helical asymmetry occurs only through resonance bands (in momentum space) if the expansion of the universe is neglected.
This phenomenon is strikingly similar to the production of fermions and bosons during preheating.
Our results confirm the presence of resonance bands in the static universe, where our formulas are significantly simplified, as well as their disappearance when expansion is taken into account.
In the expanding unverse, the presence of the pseudo-mass $m_5$ causes deviations from Gaussian distributions, exhibiting distinct amplitudes for each helical state.
Mathematically, this departure is because of the appearance of imaginary part of first-order adiabatic phase $\Theta^{(r)}_{1}$.
On the other hand, the real part of $\Theta^{(r)}_{1}$ significantly increases the number density after 8 oscillations of inflaton due to the coherent superposition.
Numerical results confirm these two conclusions.

We emphasize that, although previous studies of fermionic preheating have typically assumed a high energy scale with $m_\phi\simeq3\times10^{13}$ GeV, this assumption is not essential in the present analysis. 
The mechanism responsible for the helical asymmetry, which is driven by the sign-changing pseudo-mass term and the imaginary contribution to the adiabatic phase $\Theta_1^{(r)}$, is independent of the inflaton mass $m_\phi$. 
Consequently, present results are also valid to the lower-scale preheating scenarios, with extending the potential implications for leptogenesis and baryogenesis adapted to the inflationary models. 
In principle, this approach can be extended to low-energy inflationary models, such as small-field and hybrid inflation~\cite{hybrid}. 
It is also applicable to the generation of helical magnetic fields during preheating, such as in low-scale electroweak hybrid inflation models \cite{EW} and in gauge preheating models \cite{chiralGW2}. 

The analytical and numerical framework developed in this work can be readily extended to more general symmetry-breaking scenarios. 
For instance, variations in the form or strength of the helical asymmetry, such as momentum-dependent pseudoscalar couplings \cite{pseudoscalar6}, QCD theories \cite{QCD1,QCD2}, or electroweak interactions \cite{EW}, can be systematically incorporated. 
Similarly, the inclusion of additional scalar fields, vector fields, or alternative symmetry-breaking structures (e.g., chiral or axial couplings) 
can be analyzed within the same WKB approach by appropriately modifying the background evolution and the effective pseudo-mass term.
Such generalizations are particularly available for exploring the phenomenological implications of helical fermion production. 
They open new sight for studying leptogenesis and baryogenesis mechanisms driven by helical asymmetries, as well as potential connections to dark matter production \cite{ADM1,ADM2}, 
gravitational wave signals \cite{chiralGW1,chiralGW2}, and other cosmological observables \cite{other}. 
Moreover, the framework is naturally compatible with a broad range of inflationary energy scales, from high-scale models to low-energy scenarios such as small-field or hybrid inflation, 
thereby providing a versatile tool for investigating helical asymmetries across diverse early-universe environments.

\begin{acknowledgements}
This work was supported by the National Natural Science Foundation of China (Grant No. 12275143), Central Guidance for Local Science and Technology Development Fund Project (Grands No. 2024ZY0113, 2025ZY0020),
Inner Mongolia Natural Science Foundation (Grants No. 2024SHZR0009, 2026LHMS0083).
\end{acknowledgements}

\end{document}